\documentclass[reprint,amsmath,amssymb,aps,physrev,superscriptaddress]{revtex4-2}
\usepackage{color,graphicx}
\usepackage{array,amsmath,amssymb,pifont,mathrsfs,latexsym}
\usepackage{enumitem,multirow}
\usepackage{hyperref}
\hypersetup{
    pdfauthor={kin},
    bookmarksopen=false,
    colorlinks=true,
    linkcolor=blue,
    filecolor=magenta,      
    urlcolor=cyan,
    pdfpagemode=empty,
    pdfstartview={FitH}
}
\usepackage[T1]{fontenc}
\usepackage[utf8]{inputenc}
\usepackage{booktabs}
\usepackage{physics}
\usepackage{siunitx}
\usepackage{multirow}
\usepackage{ulem}
\usepackage{bm}
\usepackage{makecell}
\usepackage{array}
\graphicspath{{figures/}}

\begin{document}


\title{Impact of the in-medium cross section on cluster spectra in ${}^{40,48}\mathrm{Ca}+{}^{58,64}\mathrm{Ni}$ collisions at $56$ and $140$ $\mathbf{\mathrm{MeV}}/\mathrm{\mathbf{nucleon}}$}

\author{C.K.~Tam}
\affiliation{Department of Physics, Western Michigan University, Kalamazoo, Michigan 49008, USA}
\author{Z.~Chaj\k{e}cki}
\email{Contact author : zbigniew.chajecki@wmich.edu}
\affiliation{Department of Physics, Western Michigan University, Kalamazoo, Michigan 49008, USA}

\author{R.S.~Wang}
\affiliation{Facility for Rare Isotope Beams, Michigan State University, East Lansing, MI, 48824, USA}

\author{F.C.E.~Teh}
\affiliation{Facility for Rare Isotope Beams, Michigan State University, East Lansing, MI, 48824, USA}
\affiliation{Department of Physics and Astronomy, Michigan State University, East Lansing, Michigan 48824, USA}
\affiliation{Carnegie Mellon University, Pittsburgh, Pennsylvania 15213, USA}

\author{N.~Ikeno}
\affiliation{Department of Regional Environment, Tottori University, Tottori 680-8551, Japan}

\author{W.G.~Lynch}
\affiliation{Facility for Rare Isotope Beams, Michigan State University, East Lansing, MI, 48824, USA}
\affiliation{Department of Physics and Astronomy, Michigan State University, East Lansing, Michigan 48824, USA}

\author{A.~Ono}
\affiliation{Department of Physics, Graduate School of Science, Tohoku University, Sendai 980-8578, Japan}

\author{M.B.~Tsang}
\affiliation{Facility for Rare Isotope Beams, Michigan State University, East Lansing, MI, 48824, USA}
\affiliation{Department of Physics and Astronomy, Michigan State University, East Lansing, Michigan 48824, USA}

\author{A.~Anthony}
\affiliation{Facility for Rare Isotope Beams, Michigan State University, East Lansing, MI, 48824, USA}
\affiliation{Department of Physics and Astronomy, Michigan State University, East Lansing, Michigan 48824, USA}
\affiliation{Department of Physics and Astronomy, High Point University, High Point, NC, 27268, USA}

\author{S.~Barlini}
\affiliation{Dipartimento di Fisica, Università degli Studi di Firenze, 50019 Firenze, Italy}
\affiliation{Istituto Nazionale di Fisica Nucleare, Sezione di Firenze, 50019 Firenze, Italy}

\author{J.~Barney}
\affiliation{Facility for Rare Isotope Beams, Michigan State University, East Lansing, MI, 48824, USA}
\affiliation{Department of Physics and Astronomy, Michigan State University, East Lansing, Michigan 48824, USA}
\affiliation{Los Alamos National Laboratory, Los Alamos, NM 87545, USA}

\author{K.W.~Brown}
\affiliation{Facility for Rare Isotope Beams, Michigan State University, East Lansing, MI, 48824, USA}
\affiliation{Department of Chemistry, Michigan State University, East Lansing, Michigan 48824, USA}

\author{A.~Camaiani}
\affiliation{Dipartimento di Fisica, Università degli Studi di Firenze, 50019 Firenze, Italy}
\affiliation{Istituto Nazionale di Fisica Nucleare, Sezione di Firenze, 50019 Firenze, Italy}

\author{A.~Chbihi}
\affiliation{Grand Accélérateur National d'lons Lourds, Bld Henri Becquerel, BP 5027, 14021 Caen Cedex, France}

\author{D.~Dell'Aquila}
\affiliation{Facility for Rare Isotope Beams, Michigan State University, East Lansing, MI, 48824, USA}
\affiliation{Dipartimento di Fisica “E. Pancini”, Università degli Studi di Napoli “Federico II”, Via Cintia, 80126, Napoli, Italy}
\affiliation{Istituto Nazionale di Fisica Nucleare, Sezione di Napoli, Via Cintia, 80126, Napoli, Italy}

\author{J.~Estee}
\affiliation{Facility for Rare Isotope Beams, Michigan State University, East Lansing, MI, 48824, USA}
\affiliation{Department of Physics and Astronomy, Michigan State University, East Lansing, Michigan 48824, USA}

\author{A.~Galindo-Uribarri}
\affiliation{Physics Division, Oak Ridge National Laboratory, Oak Ridge, Tennessee 37831, USA}

\author{F.~Guan}
\affiliation{Department of Physics, Tsinghua University, Beijing 100084, China}

\author{B.~Hong}
\affiliation{Department of Physics, Korea University, Seoul 02841, Korea}

\author{T.~Isobe}
\affiliation{RIKEN Nishina Center, Hirosawa 2-1, Wako, Saitama 351-0198, Japan}

\author{G.~Jhang}
\affiliation{Facility for Rare Isotope Beams, Michigan State University, East Lansing, MI, 48824, USA}

\author{O.B.~Khanal}
\affiliation{Department of Physics, Western Michigan University, Kalamazoo, Michigan 49008, USA}
\affiliation{National Cancer Institute, National Institute of Health, Bethesda, MD,20892,USA}

\author{Y.J.~Kim}
\affiliation{Institute for Basic Science, Daejeon, Korea, 34126}

\author{H.S.~Lee}
\affiliation{Institute for Basic Science, Daejeon, Korea, 34126}

\author{J.W.~Lee}
\thanks{Jong-Won Lee}
\affiliation{Department of Physics, Korea University, Seoul 02841, Korea}

\author{J.-W.~Lee}
\thanks{Jung-Woo Lee}
\affiliation{Department of Physics, Korea University, Seoul 02841, Korea}

\author{J.~Manfredi}
\affiliation{Facility for Rare Isotope Beams, Michigan State University, East Lansing, MI, 48824, USA}
\affiliation{Department of Physics and Astronomy, Michigan State University, East Lansing, Michigan 48824, USA}

\author{L.~Morelli}
\affiliation{Dipartimento di Fisica, Università degli Studi di Bologna, 40126 Bologna, Italy}
\affiliation{Grand Accélérateur National d'Ions Lourds (GANIL), CEA/DRF-CNRS/IN2P3, Bd. Henri Becquerel, 14076 Caen, France}

\author{P.~Morfouace}
\affiliation{Facility for Rare Isotope Beams, Michigan State University, East Lansing, MI, 48824, USA}
\affiliation{CEA, DAM, DIF, F-91297 Arpajon, France}


\author{S.H.~Nam}
\affiliation{Department of Physics, Korea University, Seoul 02841, Korea}

\author{C.Y.~Niu}
\affiliation{Facility for Rare Isotope Beams, Michigan State University, East Lansing, MI, 48824, USA}

\author{E.~Padilla-Rodal}
\affiliation{Instituto de Ciencias Nucleares, UNAM, 04510, Mexico, D.F., Mexico}

\author{J.~Park}
\affiliation{Department of Physics, Korea University, Seoul 02841, Korea}



\author{S.~Sweany}
\affiliation{Facility for Rare Isotope Beams, Michigan State University, East Lansing, MI, 48824, USA}
\affiliation{Department of Physics and Astronomy, Michigan State University, East Lansing, Michigan 48824, USA}

\author{C.Y.~Tsang}
\affiliation{Facility for Rare Isotope Beams, Michigan State University, East Lansing, MI, 48824, USA}
\affiliation{Department of Physics and Astronomy, Michigan State University, East Lansing, Michigan 48824, USA}
\affiliation{Kent State University, 800 E Summit St, Kent, OH, 44240, USA}

\author{G.~Verde}
\affiliation{Istituto Nazionale di Fisica Nucleare-Sezione di Catania, 95123 Catania, Italy}

\author{J.~Wieske}
\affiliation{Facility for Rare Isotope Beams, Michigan State University, East Lansing, MI, 48824, USA}
\affiliation{Department of Physics and Astronomy, Michigan State University, East Lansing, Michigan 48824, USA}

\author{K.~Zhu}
\affiliation{Facility for Rare Isotope Beams, Michigan State University, East Lansing, MI, 48824, USA}
\affiliation{Department of Physics and Astronomy, Michigan State University, East Lansing, Michigan 48824, USA}

\date{\today}

\begin{abstract}
Although significant efforts have been made to investigate the density dependence of the nuclear symmetry energy, the influence of the in-medium cross section on particle production in transport models is not well constrained. The in-medium cross section reflects the dynamic situation of the medium such as a nontrivial phase space distribution. In this study, we analyze the transverse momentum spectra of $p$, $d$, $t$, ${}^3{\mathrm{He}}$ and $\alpha$ particles emitted near mid-rapidity in central $^{40,48}\mathrm{Ca}$ + $^{58, 64}\mathrm{Ni}$ reactions at $56$ and $140$ $\mathrm{MeV}/\mathrm{nucleon}$. The Antisymmetrized Molecular Dynamics ($\mathrm{AMD}$) model is chosen as the transport model for data comparison. Central events are selected based on charged-particle multiplicity in both the experimental data and AMD calculations after applying an experimental filter. Our results show that the in-medium nucleon-nucleon scattering cross-sections are more strongly reduced at $56$ $\mathrm{MeV}/\mathrm{nucleon}$ than at $140$ $\mathrm{MeV}/\mathrm{nucleon}$ incident energy.
\end{abstract}

\maketitle

\section{Introduction}
Constraining the symmetry energy term in the nuclear equation of state (EoS) is of central importance in the field of nuclear physics and astrophysics. The symmetry-energy potential affects the binding energy of a system to favor the formation of a more isospin-symmetric nucleus. This behavior opposes the trend of the Coulomb interaction, which favors the formation of neutron-rich nuclei. The interplay of these two potential terms limits the isospin asymmetry, $\delta = (N-Z) / A $, of nuclei. In astrophysics, the EoS serves as input to the Tolman-Oppenheimer-Volkoff (TOV) equation~\cite{Oppenheimer:1939ne,Tolman:1939jz} for which the solution determines the structure of static compact astronomical objects. Particularly in neutron stars, the density dependence of the symmetry energy ($\mathcal{E}_{sym} = S(\rho)\delta^2$) significantly impacts the pressure-density relation, which determines various bulk properties such as internal structure, mass-radius relations, and cooling mechanisms~\cite{Lattimer:2004pg, Lattimer:2016}. 
    
Various efforts have been made in astronomical observations~\cite{Doroshenko:2022nwp} to constrain the EoS of cold dense matter. In particular, the tidal deformability extracted from the recent gravitational wave observations~\cite{LIGOScientific:2017vwq} and the mass-radius correlations from the NICER collaboration~\cite{Riley:2019, Miller:2019} have placed important constraints on the total pressure of neutron stars at supra-saturation density. However, large uncertainties for the symmetry energy~\cite{Tsang:2020lmb, Tsang:2019mlz} remain near twice saturation density due to the lack of good quality ta~\cite{Tsang:2023vhh}.

Heavy-ion collisions (HICs) provide a terrestrial means to create nuclear mediums at a wide range of densities. At intermediate energies, particle production in HICs is a consequence of dynamical processes in the early stage ($\sim 10^{-22}$ s) of the reaction when the projectile and target nuclei overlap, followed by statistical decay processes which extend to a much longer time scale. Depending on the incident energy and the impact parameter, the system can be compressed to around twice the saturation density. Violent collisions among the constituent nucleons cause the system to thermalize~\cite{Ono:2019jxm}. In this dense, hot collision center, the dynamics of the nucleons and light clusters are significantly affected by the symmetry energy~\cite{Chen:2004kj}. Light clusters such as deuterons, tritons, and $\alpha$ particles are continuously formed and disintegrated. Depending on density and momentum, many clusters can co-exist and contribute to the dynamics of the system~\cite{Ono:2016xun}. The compressed system then expands gradually until it reaches freeze-out, where particles stop interacting with each other.

Transport theory has been the main model to simulate nucleus-nucleus collisions and extract physics from the dynamical processes~\cite{TMEP:2022xjg}. In this framework, the equation of motion of nucleons in a nuclear medium evolves under the effect of a self-consistent mean-field potential and the two-body scattering. Important physics, such as the density and isospin dependence of the symmetry energy, can be inferred by comparing observables from experimental data to transport model predictions with different prescribed mean-field potential parameters. For instance, a double neutron-proton yield ratio has been proposed as a sensitive probe to the density dependence of $S(\rho)$~\cite{Coupland:2014gya, Morfouace:2019jky}.

In nuclear collisions below 1 $\mathrm{GeV}/\mathrm{nucleon}$, a significant portion of nucleons are emitted in the form of light clusters, which indicates that cluster correlations should be explicitly considered in the scattering processes. Unlike most transport models where the cluster is only recognized at certain freeze-out configurations, $\mathrm{AMD}$ directly incorporates cluster correlations in the final state of two-nucleon collisions during the dynamical evolution of the system~\cite{Ono:2020zsx, Ikeno:2016xpr}. After the freeze-out, the excited fragments will undergo sequential decay, which leads to a final state composed of light particles and fragments~\cite{Tan:2003bi}. These de-excited fragments are compared to data. $\mathrm{AMD}$ has successfully reproduced particle spectra~\cite{SpiRIT:2021och, SpiRIT:2022sqt, INDRA:2023myq}, spectral ratios~\cite{Ono:2016xun}, charge distributions~\cite{Ono:2019jxm,Ono:2020zsx} and stopping observables such as the portion of energy transferred from projectile to target in the transverse direction~\cite{Ono:2020zsx} observed in various reactions.
    
In transport models, the nucleon-nucleon scattering cross-section $\sigma_{\mathrm{NN}}$ accounts for the stochastic nucleon-nucleon collision  in a nuclear medium. It is well established that the magnitude of $\sigma_{\mathrm{NN}}$ is reduced compared to that in free space~\cite{Danielewicz:2002he, Barker:2016hqv, FOPI:2010xrt, INDRA:2002kji}. However, the reduction is not well constrained, and different prescriptions have been used to describe reactions at different densities and energies~\cite{FOPI:2010xrt, SpiRIT:2023htl}. 
In this work, $\mathrm{AMD}$ calculations with different reductions in $\tilde{\sigma}_\mathrm{NN}$ are used to describe transverse momentum spectra of light-charged particles from $^{40,48}$Ca + $^{58, 64}$Ni reaction at incident energies of $56$ and $140$ $\mathrm{MeV}/\mathrm{nucleon}$. A direct comparison between data and model calculations provides information about the dependence of the in-medium effect on reaction conditions. To avoid confusion with the free nucleon-nucleon cross-section, we adopt the symbol $\tilde{\sigma}_\mathrm{NN}$ to indicate the modified $\sigma_\mathrm{NN}$ used in our work.

\begin{table}[h]
    \centering
    \small
    \begin{tabular}{@{}c c c c c c@{}}
        \toprule
        \textbf{Beam} & \textbf{Target} & {$(N-Z)/A$} & \makecell{\textbf{Beam Energy} \\ (MeV/nucleon)} & $\boldsymbol{\beta_\mathrm{cms}}$ & \makecell{\textbf{Beam} \\ \textbf{Rapidity}} \\
        \midrule
        ${}^{40}\mathrm{Ca}$ & ${}^{58}\mathrm{Ni}$ & 0.020 & \makecell{56 \\ 140} & \makecell{0.1392 \\ 0.2171} & \makecell{0.3410 \\ 0.5354} \\
        ${}^{40}\mathrm{Ca}$ & ${}^{64}\mathrm{Ni}$ & 0.077 & \makecell{56 \\ 140} & \makecell{0.1315 \\ 0.2055} & \makecell{0.3410 \\ 0.5354} \\
        ${}^{48}\mathrm{Ca}$ & ${}^{58}\mathrm{Ni}$ & 0.094 & \makecell{56 \\ 140} & \makecell{0.1540 \\ 0.2393} & \makecell{0.3417 \\ 0.5364} \\
        ${}^{48}\mathrm{Ca}$ & ${}^{64}\mathrm{Ni}$ & 0.143 & \makecell{56 \\ 140} & \makecell{0.1461 \\ 0.2274} & \makecell{0.3417 \\ 0.5364} \\
        \bottomrule
    \end{tabular}
    \caption{Characteristics of the four reactions studied in this work. $\beta_{cms}$ is the velocity of the center of mass frame in units of speed of light. $y_{\mathrm{beam}}$ is rapidity of beam nuclei in the lab frame. $\delta=(N-Z)/A$ is the isospin asymmetry of the system.}
    \label{tab:reactions}
\end{table}

\section{Experiment}
    \label{sec:experiment}
    The experiment was conducted at the National Superconducting Cyclotron Laboratory (NSCL) at Michigan State University. Primary beams of $^{40,48}\mathrm{Ca}$ at $56$, $140$ $\mathrm{MeV}/\mathrm{nucleon}$ were impinged on targets of 5.0 and 5.3 mg/cm$^2$ of $^{58,64}$Ni respectively. Table~\ref{tab:reactions} gives a summary of the characteristics of the studied systems. The detection system consists of three parts: the impact parameter detection system (Microball)~\cite{Sarantites:1996}, the charged-particle detection system (HiRA10)~\cite{Wallace:2007,DellAquila:2019rat}, and the neutron detection system with charged particle vetos~\cite{zecher:1997,Zhu:2020}. Details of the experiment setup can be found in Ref.~\cite{Paneru:2023, Sweany:2020abc, Teh:2023szo}. In this study, we mainly focus on the analysis of charged particles from central collisions detected in HiRA10 in coincidence with particles detected in the Microball.

    \begin{figure}
        \includegraphics[width=0.475\textwidth]{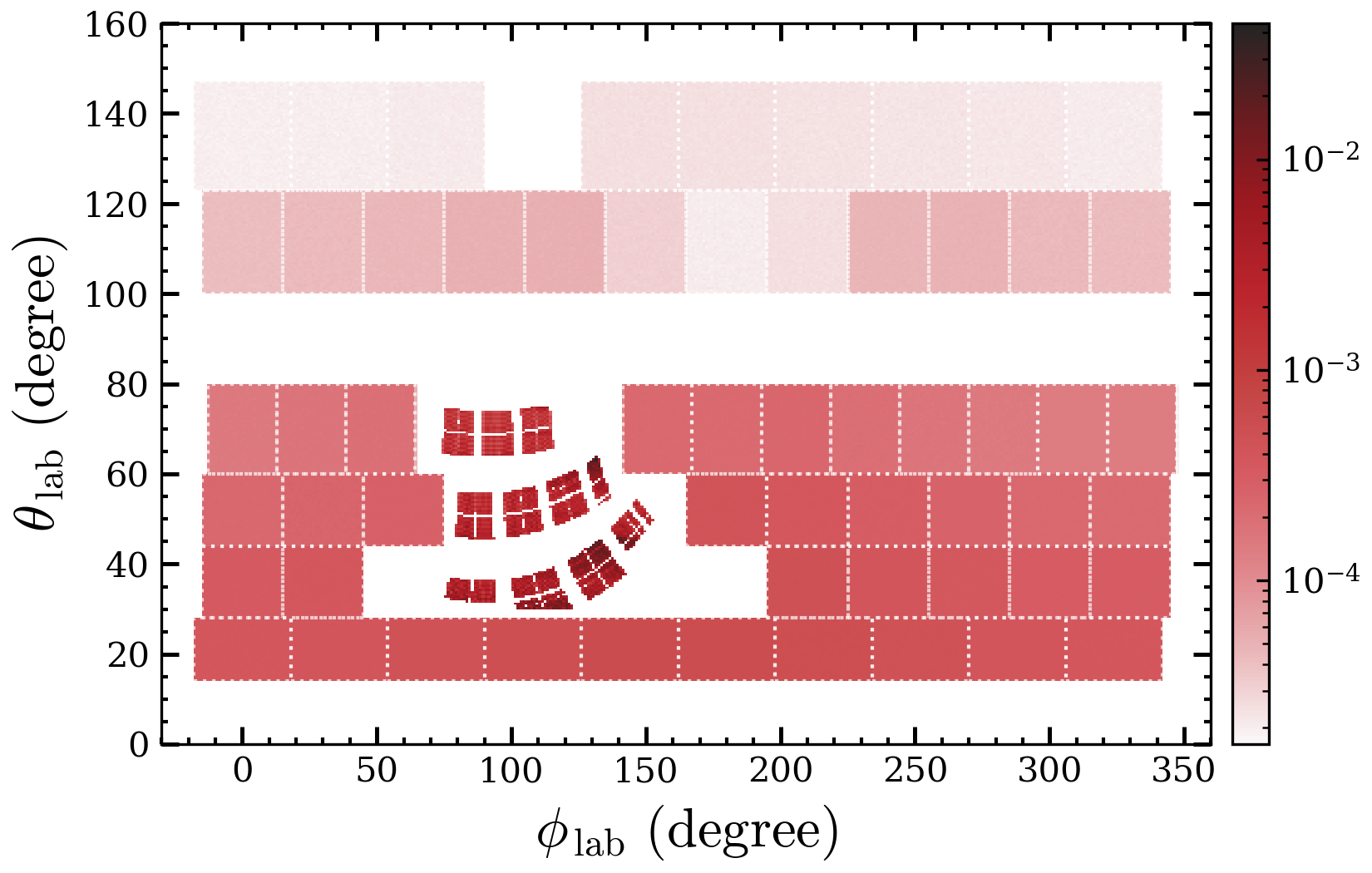}
        \caption{
        Charged-particle hit pattern of the Microball and HiRA10 detectors. The boundaries of each rectangular block represent CsI(Tl) crystals in Microball. The coverage at $\theta_\mathrm{lab} \in (30^\circ, 75^\circ)$ corresponds to charged particles detected by HiRA10.}
        \label{fig:1x1-hit-pattern}
    \end{figure}

    The target was placed inside the Washington University Microball, a nearly $4\pi$ detector packed with CsI(Tl) crystals arranged in rings at different polar angles. To avoid beam particles from directly hitting Microball, CsI(Tl) crystals at $\theta_{\mathrm{lab}} < 18^\circ$ and $\theta_{\mathrm{lab}} > 147^\circ$ were removed. A total of 10 crystals at $\theta_{\mathrm{lab}}\in (28^\circ,80^\circ)$ were removed to allow the passage of charged particles to the HiRA10 detector. Finally, detectors at $\theta_{\mathrm{lab}}\in (80^\circ,100^\circ)$ were removed for the entire azimuthal angle for the installation of the target ladder. One detector at $\theta_{\mathrm{lab}}\in (123^\circ,147^\circ)$, $\phi\in(90^\circ, 126^\circ)$ was excluded in the analysis due to bad performance. This left 59 crystals covering about $65$\% of the solid angle, see Fig.~\ref{fig:1x1-hit-pattern}.

    
    The impact parameter of each event can be estimated from the multiplicity of particles, $N_C$, that hit the Microball. A detailed description of this method can be found in~\cite{Sweany:2020abc}. During the experiments, only events with Microball multiplicity greater than 5 were recorded. To select central events with $b\lesssim 3$ fm, a software gate on $N_C\ge 10$ for reactions at $56$ $\mathrm{MeV}/\mathrm{nucleon}$ and $N_C\ge 14$ at $140$ $\mathrm{MeV}/\mathrm{nucleon}$ was applied in the analysis. The detailed procedure will be presented in Sec.~\ref{sec:event-selection}.
    
    The upgraded High-Resolution Array (HiRA10) was positioned $\sim 33$ $\mathrm{cm}$ from the target on the right-hand side when viewed from the beam. It is comprised of 12 independent $\Delta E$--$E$ telescopes, which can be arranged into various configurations to accommodate specific experimental requirements. In this work, HiRA10 was arranged into 3 separate towers of 4 telescopes each to span $\theta_{\mathrm{lab}}\in(30^\circ,75^\circ)$, focusing on the detection of light-charged particles at mid-rapidity. The phase space occupied by the microball scintillators (squares) and HiRA10 detectors (shaded squares inside the large opening of the microball) are shown in Fig. 1. 
    
    In each telescope, a $1.5\,\mathrm{mm}$ thick Double-Sided Silicon Strip Detector (DSSD) serves as the $\Delta E$ detector. Each of the front and back sides of the DSSD is composed of 32 strips with a spacing of $1.95\,\mathrm{mm}$. The strips in the front and back sides are oriented perpendicularly to each other, creating a measurement grid with an angular resolution better than $0.5^\circ$~\cite{DellAquila:2019rat}. Followed by the DSSD is a composite of four $10\,\mathrm{cm}$-thick CsI(Tl) crystals (E-detector), a major upgrade from the previous $4\,\mathrm{cm}$ design~\cite{Wallace:2007}. The upgrade allows charged particles with $E_{\mathrm{lab}}<100$ $\mathrm{MeV}/\mathrm{nucleon}$ to completely deposit their energies. We identify $p$, $d$, $t$, $^3{\mathrm{He}}$ and $\alpha$ particles with high statistics. For simplicity, CsI(Tl) will be referred to as CsI in the rest of this manuscript. Details in the calibrations and performance of HiRA10 can be found in~\cite{DellAquila:2019rat, Sweany:2021hjd}.

\section{Transport model Simulations} 
\begin{figure}
    \includegraphics[width=0.4\textwidth]{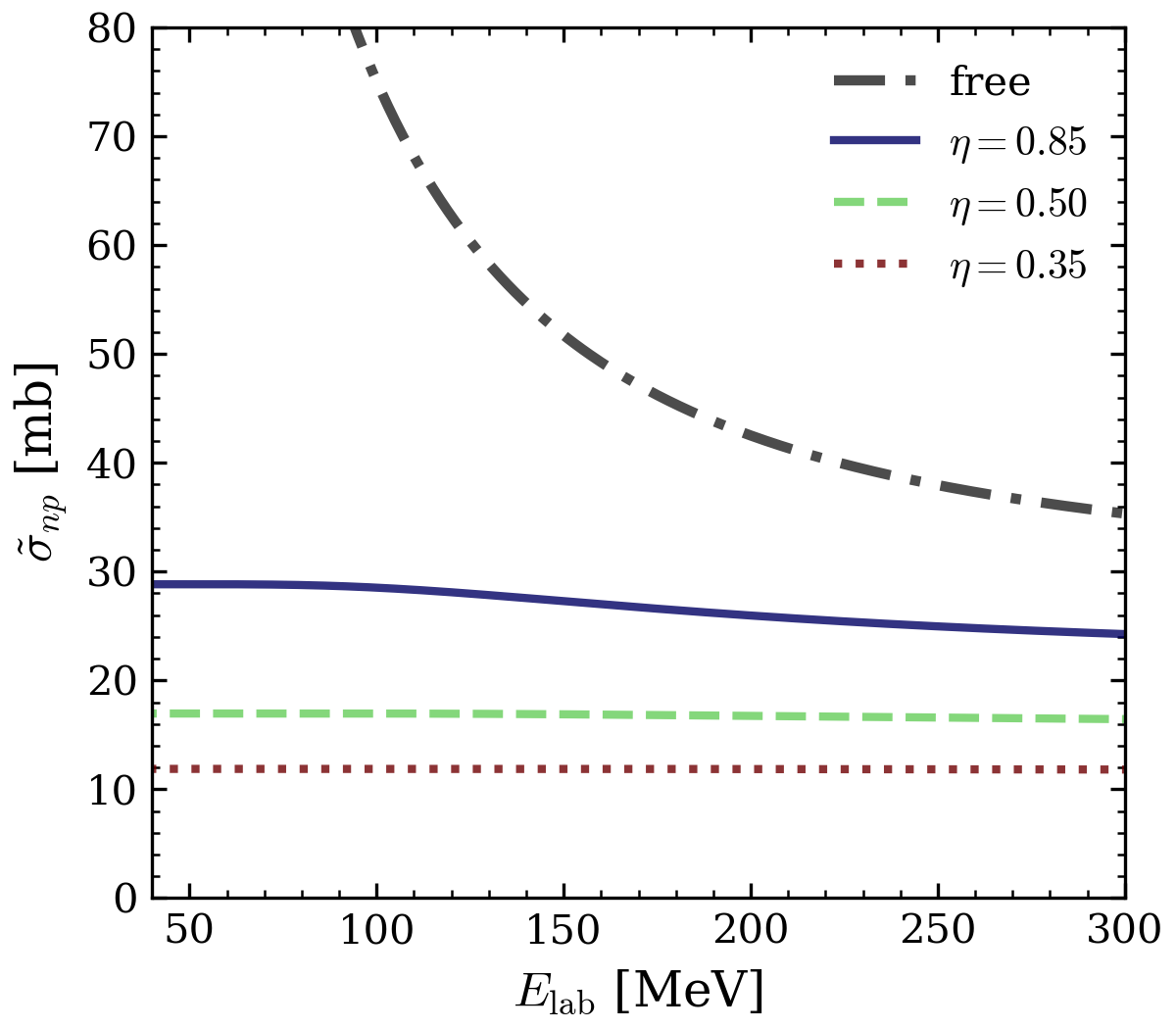}
    \caption{Effect of the screening parameter $\eta$ on $\tilde{\sigma}_{np}$ at saturation density. This $\tilde{\sigma}_{np}$ is used to obtain the matrix element $|M|^2$ in medium, which then determines the $NN$ cross section by Eq.~\eqref{amd_coll_sigma}.
    } 
    \label{fig:1x1-screened-cross-section-eta}
\end{figure}
\subsection{Antisymmetrized Molecular Dynamics}
    \label{sec:amd}
    The Antisymmetrized Molecular Dynamics (AMD) model~\cite{Ono:2020zsx,SpiRIT:2020sfn,SpiRIT:2021och,SpiRIT:2022sqt,TMEP:2022xjg} is used to simulate the dynamic phase of the reaction systems up to $300$ $\mathrm{fm}/c$ in reactions at $140$ $\mathrm{MeV}/\mathrm{nucleon}$, and up to $500$ $\mathrm{fm}/c$ in reactions at $56$ $\mathrm{MeV}/\mathrm{nucleon}$. The average emission times for low-momentum $\alpha$ particles with $p_{\mathrm{T}}/\mathrm{A} < 200$ $\mathrm{MeV}/c$ are slightly below $200$ and $120$ $\mathrm{fm}/c$ in ${}^{48}\mathrm{Ca} + {}^{64}\mathrm{Ni}$ at $56$ and $140\,\mathrm{MeV}/\mathrm{nucleon}$, respectively. This justifies the selected time limit when the dynamical phase ends and the excited fragments undergo statistical decay~\cite{Puhlhofer:1977zz, Maruyama:1992bh}. The impact parameter ranges from $0$ to $8$ $\mathrm{fm}$ following a geometric distribution. The SLy4 effective interaction~\cite{Chabanat:1997,Chabanat:1997un} was employed in this work. It corresponds to the symmetry energy slope parameter $L=46\,\mathrm{MeV}$ and exhibits a quadratic momentum dependence of the mean field, characterized by an effective mass $m^*/m=0.70$ in symmetric matter at saturation density. 

    The ability to describe cluster formation is one of the distinctive features of $\mathrm{AMD}$ model, compared to other transport model approaches~\cite{TMEP:2022xjg,Ono:2019jxm}. Cluster correlation is implemented in the final state of each two-nucleon collision ($N_1 + N_2 + B_1 + B_2 \rightarrow C_1 + C_2$), where $N_1$ and $N_2$ are the colliding nucleons. The process includes the special cases where a surrounding particle $B_1$ and/or $B_2$ might be empty~\cite{Ono:2020zsx, Ikeno:2016xpr}. For a specific configuration ($C_1, C_2$), the collision cross section is written as 
    \begin{gather}
        \label{amd_coll_sigma}
        \dfrac{d\sigma(C_1, C_2)}{d\Omega} = P(C_1, C_2, p_f, \Omega)\dfrac{p_i}{v_i}\frac{p_f}{v_f}|M|^2 \dfrac{p_f}{p_i}
    \end{gather}
    where $p_i$ and $v_i$ refer to the initial relative momentum and velocity between the scattered nucleons.
    The relative momentum vector between the nucleons after momentum transfer is denoted as $(p_f, \Omega)$, where $p_f$ is determined from the energy conservation for the adopted effective interaction. This implies that the velocity after momentum transfer $v_f = \partial E/\partial p_f$, is also dependent on the effective interaction. Any modification to the momentum-dependence of the mean field in Eq. \eqref{amd_coll_sigma} will be reflected in the factor of $p_i/v_i$ and $p_f/v_f$. The overlap probability factor for cluster formation, $P(C_1,C_2,p_f,\Omega)$, is defined by considering the non-orthogonality between the states of different configurations~\cite{Ikeno:2016xpr,Ono_2013}. 
    
    An important ingredient to the calculation  is the matrix element for two-nucleon scattering $|M|^2$. We express it as
    \begin{align}
     |M|^2 &=(2/m_N)^2 d(\tilde{\sigma}_\mathrm{NN})/d\Omega,
     \label{eq:Msigmann}
     \end{align}
     where $\tilde{\sigma}_\mathrm{NN}$ is parameterized~\cite{Coupland:2011px,Danielewicz:2002he} as
    \begin{align}
        \tilde{\sigma}_\mathrm{NN} &=\sigma_{0}\tanh{\frac{\sigma^{(\text{free})}_{NN}}{\sigma_{0}}}, \label{eq:screening}\\
        \sigma_{0} &=\eta(\rho')^{-2/3}.\label{Eq_SigmaNN}
    \end{align}
   In these formulas, the right-hand side of Eq.~\eqref{eq:Msigmann} is evaluated at an average of $p_{\text{i}}$ and $p_{\text{f}}$, and $\rho'$ is a density with a momentum cut as defined by Eq.~(161) in Ref.~\cite{TMEP:2022xjg}. The free parameter $\eta$ in Eq.~\eqref{Eq_SigmaNN} can be adjusted to control the degree of reduction of the nucleon-nucleon matrix element.  Figure~\ref{fig:1x1-screened-cross-section-eta} illustrates how varying $\eta$ impacts the $\tilde{\sigma}_{np}$ at saturation density. 
   
\subsection{Experimental filter}
    The sharp cutoff method~\cite{Cavata:1990gk} has been widely applied in experiments to select central events from observables such as $N_C$ and total kinetic energy, which exhibits a negative correlation with $b$. However, the stochastic nature of nucleon-nucleon collisions implies the existence of large fluctuations in such observables for a given $b$ in the transport model. This poses an ambiguity when comparing data with the prediction of the model. Alternative methods include using machine learning~\cite{Tsang:2021rku} to map $b$ from a selection of observables and the model-independent method based on {Bayes'} theorem~\cite{INDRA:2020kyj, Chen:2022kxe,INDRA:2023myq, INDRA-FAZIA:2024rxc}. In this work, we adopt the simple approach by applying the experiment condition in $\mathrm{AMD}$ to select events in the same manner as in the experiment. 

\begin{figure}
    \includegraphics[width=0.4\textwidth]{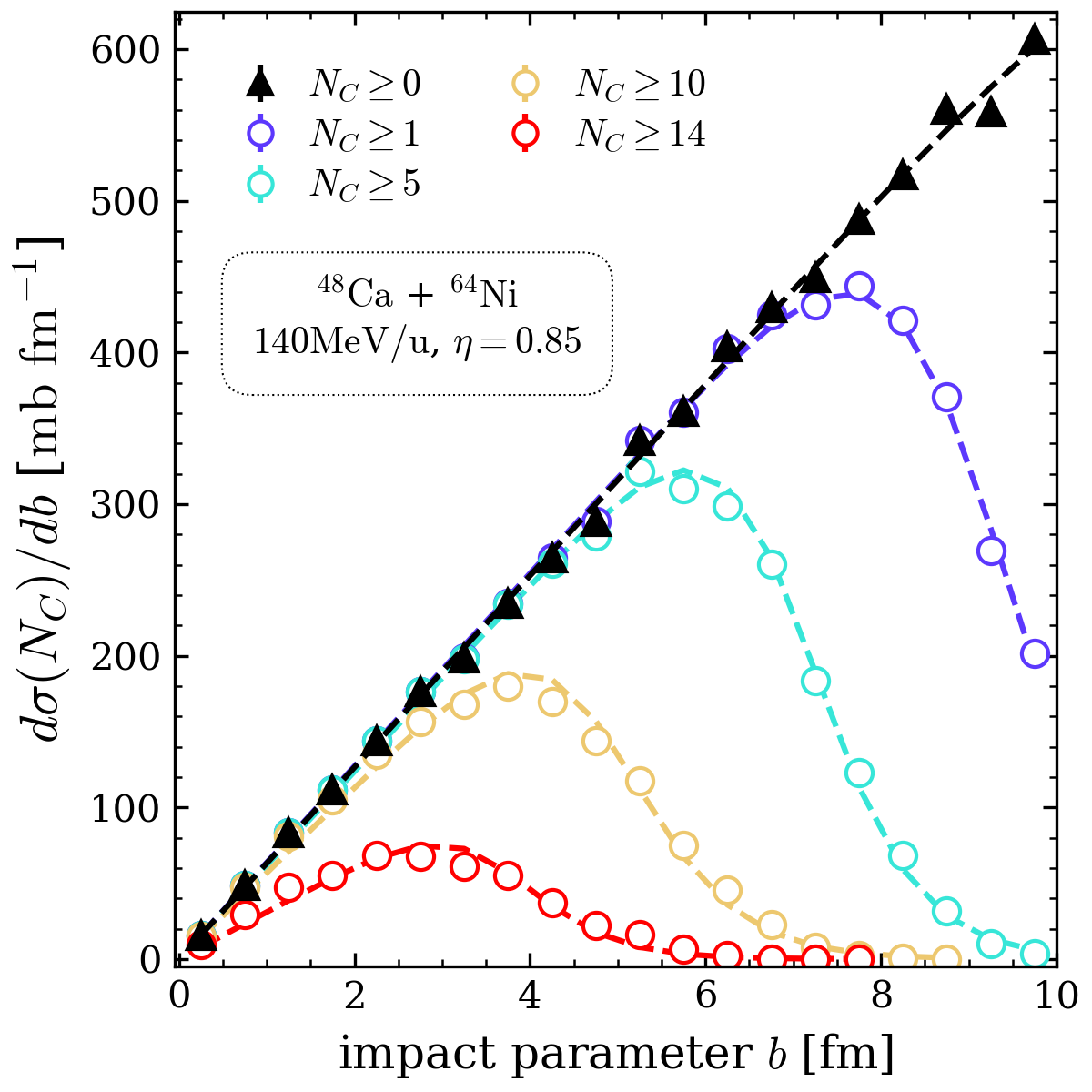}
    \caption{Differential cross section $d\sigma(N_C)/db$ from AMD calculation. The dashed lines are fits to the distribution with the equation $2\pi b / (1 + \exp((b-b_0)/\Delta b))$, see Ref.~\cite{INDRA:2020kyj}. 
    }
    \label{fig:1x1-dsigma-db}
\end{figure}

    Figure~\ref{fig:1x1-hit-pattern} shows the hit pattern of charged particles falling within the coverage of the Microball in polar ($\theta$) and azimuthal ($\phi$) angles in the laboratory frame. Each square represents the approximate coverage of a single CsI crystal in the Microball. The particle hits gradually decrease from forward to backward polar angles. In the experimental setup, the front of each CsI in the Microball was covered with a thin foil of Sn/Pb to block electrons generated in the HICs. Particles without sufficient energy to punch through the foils are not detected. Furthermore, due to the fact that a single CsI crystal can not resolve multiple signals, only a single hit is recorded in each crystal in the same event.
    Figure~\ref{fig:1x1-dsigma-db} shows that the probability of the events with large $b$ drops significantly as $N_C$ increases, consistent with~\cite{INDRA:2020kyj}.

\begin{figure}[h]
    \includegraphics[width=0.45\textwidth]{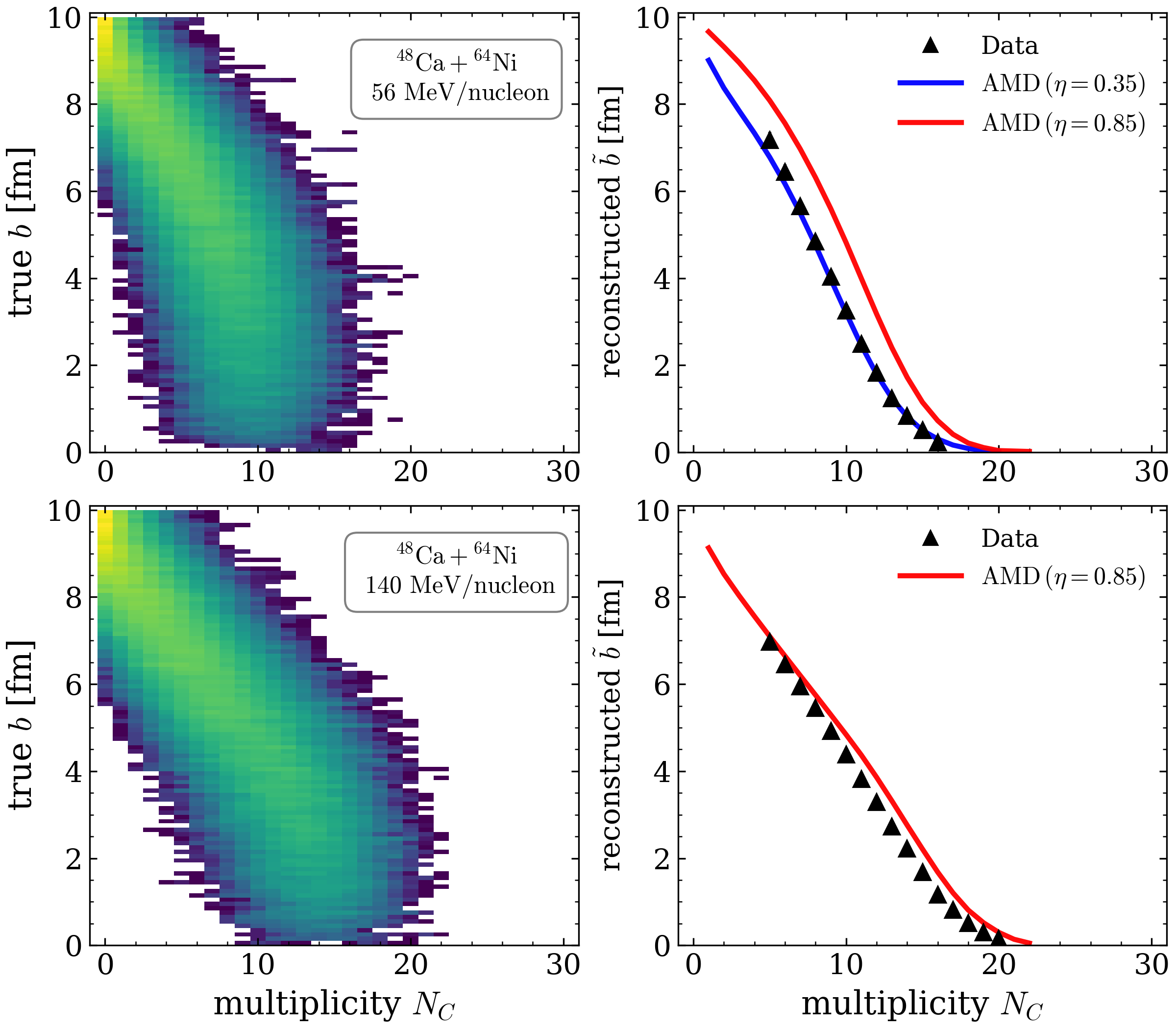}
    \caption{
        Left : Correlation between the impact parameter and the charged-particle multiplicities in $\mathrm{AMD}$ after filtering out events according to experimental gates. Right : The calculated mapping between charged-particle multiplicity $N_C$ and the estimated impact parameter in the experiment (black triangle) and $\mathrm{AMD}$ with $\eta=0.35$ (blue line) and $\eta=0.85$ (red line). The top and bottom panels refer to the reactions $^{48}$Ca + $^{64}$Ni at $56$ (top) and $140$ (bottom) $\mathrm{MeV}/\mathrm{nucleon}$, respectively.
    }
    \label{fig:amd-bmulti}
\end{figure}

\subsection{Event selection}
\label{sec:event-selection}
To make a fair comparison to experimental data, one must select events with similar characteristics in $\mathrm{AMD}$ calculations. Since the actual impact parameter cannot be directly measured, event centrality is constructed using the cross section $\sigma(N_C)$ of detecting at least $N_C$ charged particles.

In the experiment, the normalization of $\sigma(N_C)$ is determined from the total number of beam particles and the areal density of the target. For the $\mathrm{AMD}$ simulations, $\sigma(N_C)$ is calculated for the charged particle multiplicity $N_C$ after applying an experimental filter, specifically, the Microball filter described in Sec.~\ref{sec:experiment}. The left panels of Fig.~\ref{fig:amd-bmulti} show the correlation between the impact parameter and the charged-particle multiplicities in $\mathrm{AMD}$ calculations with an experimental filter.

Figure~\ref{fig:1x1-dsigma-db} shows the differential cross section $d\sigma(N_C)/db$ from AMD calculation. In $d\sigma(N_C\ge 1)/db$, peripheral events with $N_C=0$ are not included, leading to the drastic drop in $d\sigma(N_C\ge 1)/db$ beyond $b\approx 8$ $\mathrm{fm}$ as compared to the geometric distribution in raw simulation. As the minimum multiplicity increases from $N_C=1$ to $N_C=14$, the peak of $d\sigma(N_C)/db$ leans towards a smaller impact parameter. 

For convenience, experimentalists relate $N_C$ to the reconstructed impact parameter defined from simple geometry : $\sigma(N_C) = \pi \tilde{b}^2(N_C)$. The reconstructed impact parameters $\tilde{b}(N_C)$ are shown on the right-hand side of Fig.~\ref{fig:amd-bmulti}. While calculations with screening parameter $\eta=0.85$ (red) reproduce the experimental $\tilde{b}(N_C)$ at $140$ $\mathrm{MeV}/\mathrm{nucleon}$, it overestimates $\tilde{b}(N_C)$ at $56$ $\mathrm{MeV}/\mathrm{nucleon}$. To address the insufficient reduction of $|M|^2$ at $56$ $\mathrm{MeV}/\mathrm{nucleon}$, $\eta$ is adjusted to $0.35$ (blue) and successfully reproduces $\tilde{b}(N_C)$. 

With the cluster production accurately reproduced, event selection by gating on $\tilde{b}(N_C)$ or $\sigma(N_C)$ on both data and model is applied. In this study, events with $\tilde{b}(N_C)\le 3$ $\mathrm{fm}$ are selected to probe the physics from the participant zone. For reactions at $E_\mathrm{beam}=56$ $\mathrm{MeV}/\mathrm{nucleon}$, this corresponds to $N_C\ge10$ in data and $N_C\ge 10,12$ in $\mathrm{AMD}$ with $\eta=0.35,0.85$, respectively. For reactions at $E_\mathrm{beam}=140$ $\mathrm{MeV}/\mathrm{nucleon}$, it corresponds to $N_C\ge 14$ in both data and $\mathrm{AMD}$.

\section{Results and discussion}

\begin{figure}
    \includegraphics[width=0.475\textwidth]{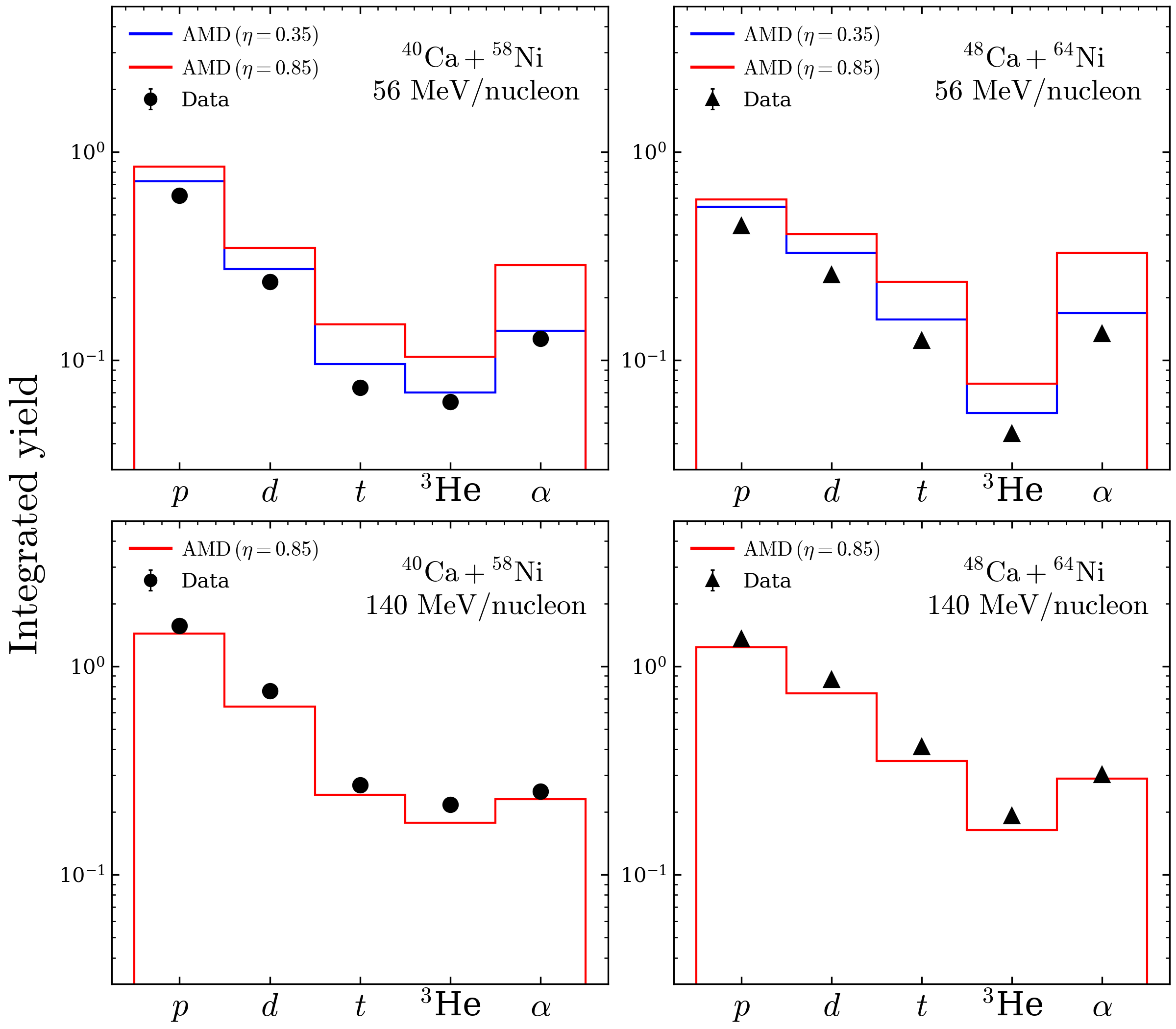}
    \caption{Integrated yields of $p$, $d$, $t$, ${}^3\mathrm{He}$, $\alpha$ of $^{40,48}\mathrm{Ca}+{}^{58,64}\mathrm{Ni}$ reactions at $56$ (top) and $140$ (bottom) $\mathrm{MeV}/\mathrm{nucleon}$. Data are represented by black markers and $\mathrm{AMD}$ calculations with $\eta=0.85$ and $\eta=0.35$ are represented by red and blue lines respectively. The integration is over the mid-rapidity region $0.4 < y_\mathrm{lab}/y_\mathrm{beam}<0.6$ and in the $p_\mathrm{T}/\mathrm{A}$ range showed in Fig.~\ref{fig:5x2-pt-e56} and Fig.~\ref{fig:5x2-pt-e140}. Statistical uncertainties are smaller than the data points.} 
    \label{fig:2x2-yields-log}
\end{figure}

The black points in Fig.~\ref{fig:2x2-yields-log} show the experimental yields of $p$, $d$, $t$, ${}^3\mathrm{He}$ and $\alpha$ particles gated by a mid-rapidity window $0.4 < y_{\mathrm{lab}}/y_{\mathrm{beam}} < 0.6$ from $^{40}\mathrm{Ca}+{}^{58}\mathrm{Ni}$ (circle) and $^{48}\mathrm{Ca}+{}^{64}\mathrm{Ni}$ (triangles) at $56$ (bottom) and $140$ (top) $\mathrm{MeV}/\mathrm{nucleon}$, respectively. Here, $y_{\mathrm{lab}}$ refers to the particle rapidity and $y_\mathrm{beam}$ refers to the beam nucleus rapidity in the laboratory frame, where $y_{\mathrm{beam}}\approx 0.54$ for $140$ $\mathrm{MeV}/\mathrm{nucleon}$ and $y_{\mathrm{beam}}\approx 0.34$ for $56$ $\mathrm{MeV}/\mathrm{nucleon}$.

In the reactions at $56$ $\mathrm{MeV}/\mathrm{nucleon}$, the production of proton-rich particles is more prominent in $^{40}\mathrm{Ca}+{}^{58}\mathrm{Ni}$ than in $^{48}\mathrm{Ca}+{}^{64}\mathrm{Ni}$, with $\sim 40\%$ more $p$ and ${}^3\mathrm{He}$. On the other hand, $^{48}\mathrm{Ca}+{}^{64}\mathrm{Ni}$ reaction produces nearly $70\%$ more $t$, and $\sim 10\%$ more $d$ and $\alpha$ particles than ${}^{40}\mathrm{Ca}+{}^{58}\mathrm{Ni}$, due to the abundance of neutrons in the colliding nuclei. A similar trend is observed in the reactions at $140$ $\mathrm{MeV}/\mathrm{nucleon}$. There are $\sim15\%$ more $p$ and ${}^3\mathrm{He}$ produced in $^{40}\mathrm{Ca}+{}^{58}\mathrm{Ni}$ than in $^{48}\mathrm{Ca}+{}^{64}\mathrm{Ni}$. The neutron-rich $^{48}\mathrm{Ca}+{}^{64}\mathrm{Ni}$ reaction produces $\sim50\%$ more $t$ and $\sim15\%$ more $d$ and $\alpha$ than $^{40}\mathrm{Ca}+{}^{58}\mathrm{Ni}$. 

We then compare the data at $E_\mathrm{beam}=140$ $\mathrm{MeV}/\mathrm{nucleon}$ to $\mathrm{AMD}(\eta=0.85)$. Stopping is modified by varying $\eta$. As seen in the red line in the bottom panels of Fig.~\ref{fig:2x2-yields-log}, the emission of $p$, $t$ and $\alpha$ particles in the mid-rapidity region are well reproduced in both reactions, while the yields of deuteron and $^3\mathrm{He}$ are underestimated by $\sim 10\%$.

Next, we move on to calculations for the reactions at $56$ $\mathrm{MeV}/\mathrm{nucleon}$. Using $\eta=0.85$, the yields of all particles are significantly overestimated, as seen from the red line in top panels of Fig.~\ref{fig:2x2-yields-log}. This is consistent with the overestimated impact parameters shown in top-right panel of Fig.~\ref{fig:amd-bmulti}. To mitigate the over-enhanced $\sigma(N_C)$, $\eta$ was adjusted to $0.35$. As shown by the blue lines in the upper panels of Fig.~\ref{fig:2x2-yields-log}, the production of all particle species aligns well with the experimental results, with deviations of approximately $10\%$. 



\begin{figure}
    \includegraphics[width=0.45\textwidth]{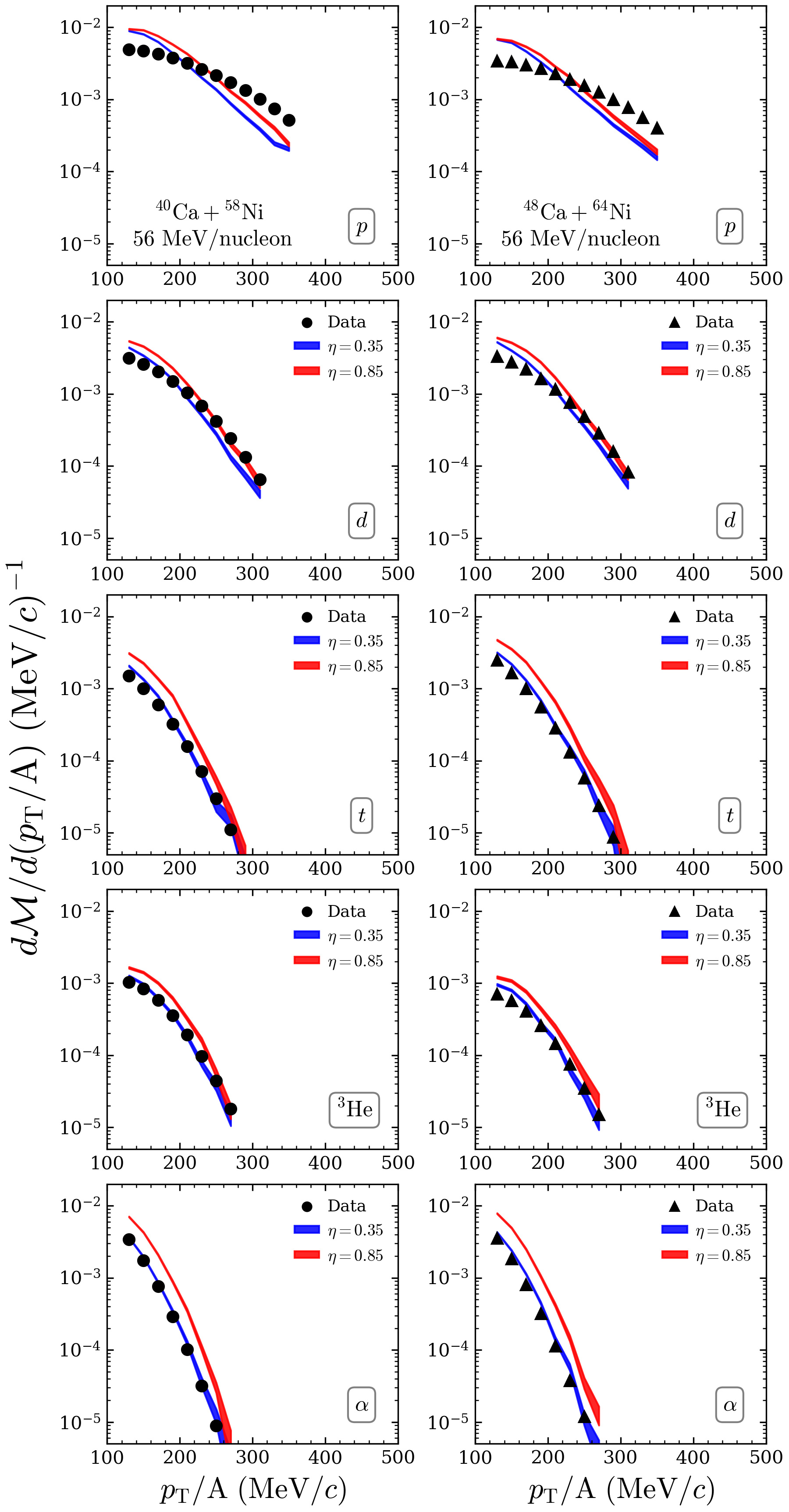}
    \caption{Transverse momentum spectra of $p$, $d$, $t$, $^3\mathrm{He}$ and $\alpha$-particle (from top to bottom) in the mid-rapidity window $0.4 \le y/y_\mathrm{beam}\le 0.6$ in $^{40}\mathrm{Ca}+^{58}\mathrm{Ni}$ (left panels) and $^{48}\mathrm{Ca}+^{64}\mathrm{Ni}$ (right panels) reactions at $56$ $\mathrm{MeV}/\mathrm{nucleon}$. Data are shown in black symbols; Calculations with $\mathrm{AMD}$ with $\eta=0.85$ and $\eta=0.35$ are shown in red and blue shades, respectively.}
    \label{fig:5x2-pt-e56}
\end{figure}
\begin{figure}
    \includegraphics[width=0.45\textwidth]{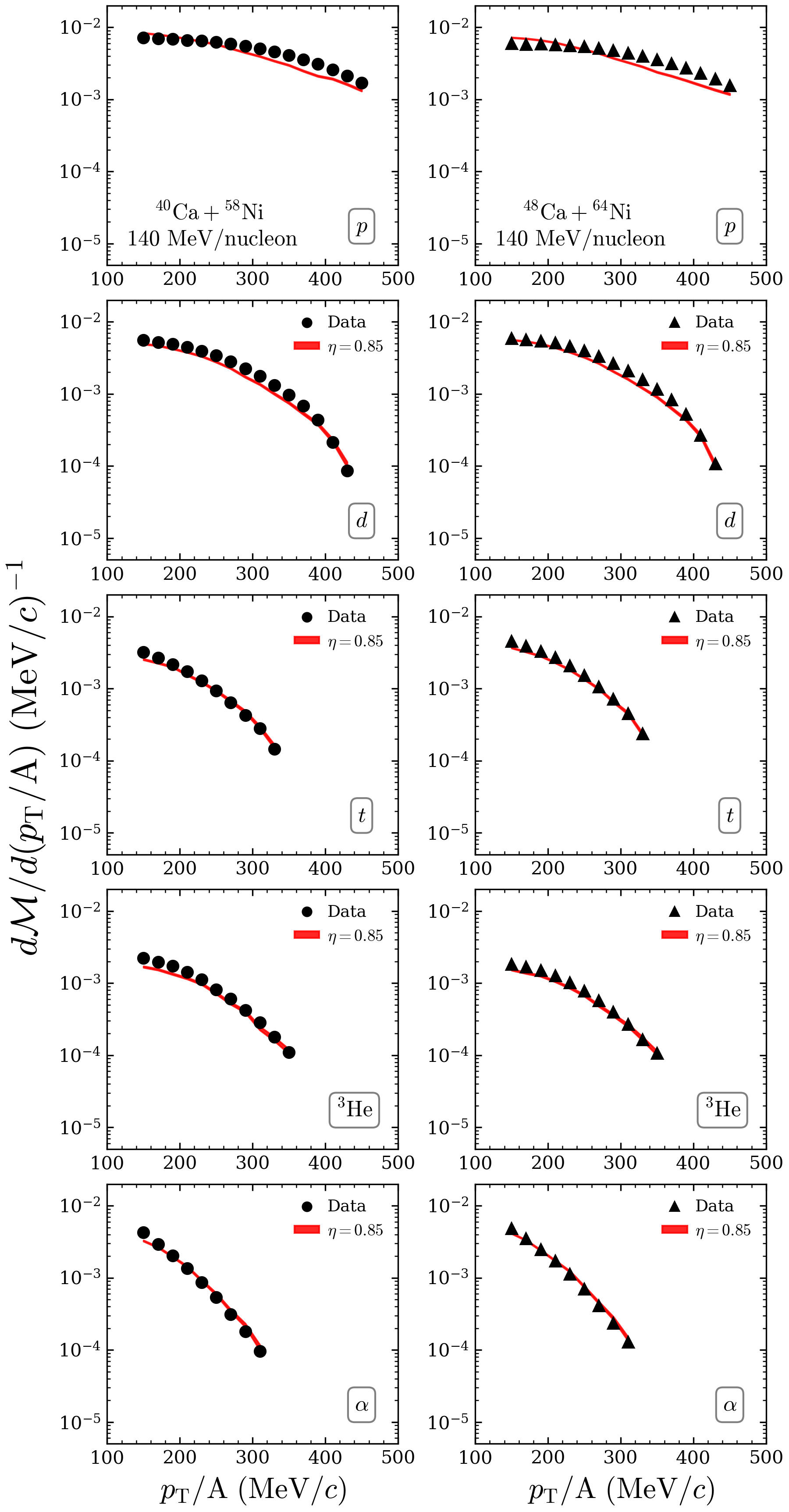}
    \caption{Transverse momentum spectra of $p$, $d$, $t$, $^3\mathrm{He}$ and $\alpha$-particle (from top to bottom) in the mid-rapidity window $0.4 \le y/y_\mathrm{beam}\le 0.6$ in $^{40}\mathrm{Ca}+^{58}\mathrm{Ni}$ (left panels) and $^{48}\mathrm{Ca}+^{64}\mathrm{Ni}$ (right panels) reactions at $140$ $\mathrm{MeV}/\mathrm{nucleon}$. Data are shown in black symbols; Calculations with $\mathrm{AMD}$ with $\eta=0.85$ are shown in red shade.}
    \label{fig:5x2-pt-e140}
\end{figure}

As a verification to the constrained in-medium two-nucleon collision parameter, the transverse momentum distribution {$d\mathcal{M}/d(p_{\mathrm{T}}/\mathrm{A})$} of light-charged particles from mid-rapidity $0.4 \le y/y_\mathrm{beam}\le 0.6$ in reactions at $56$ and $140$ $\mathrm{MeV}/\mathrm{nucleon}$ are presented in~Fig.~\ref{fig:5x2-pt-e56} and Fig.~\ref{fig:5x2-pt-e140}, respectively. Due to the geometry of HiRA10 and the energy thresholds of different particles, the coverage in $y_{\mathrm{lab}}/y_{\mathrm{beam}}$ are only complete in a window of $ 200 \lesssim p_T / \mathrm{A} \lesssim 400$ MeV$/c$. To account for the missing coverage at small $p_T / \mathrm{A}$, the spectra are scaled up by the inverse of the coverage fraction. Such a correction is valid since the spectra are approximately independent of rapidity in the region of interest.

Figure~\ref{fig:5x2-pt-e56} shows the $p_T/\mathrm{A}$ spectra for the two reactions at $56$ $\mathrm{MeV}/\mathrm{nucleon}$. The blue shaded regions represent the $\mathrm{AMD}$ calculation with $\eta=0.35$, along with the associated statistical uncertainty. The production of $d,t,{}^3\mathrm{He}$ and $\alpha$ particles is accurately reproduced over the entire acceptance of $p_T/A$. Although the integrated proton yield is reasonably reproduced, the slope of the $p$ spectrum is too steep. One possible reason is the contamination of protons emitted from the projectile-like fragments.

Figure~\ref{fig:5x2-pt-e140} shows the $p_T/\mathrm{A}$ spectra for the two reactions at $140$ $\mathrm{MeV}/\mathrm{nucleon}$. The red shaded regions represent the $\mathrm{AMD}$ calculation with $\eta=0.85$. The predictions of $d, t,{}^3\mathrm{He}$ and $\alpha$ particles are consistent with the result from the total yield. Similar to the case in $56$ $\mathrm{MeV}/\mathrm{nucleon}$, the slope of the proton spectrum is steep.

The implementation of in-medium effects is not universal across reactions at different beam energies. In our calculation, $\eta=0.35$ is consistent with data in the reaction at $56$ $\mathrm{MeV}/\mathrm{nucleon}$ and $\eta=0.85$ is consistent with data in the reaction at $140$ $\mathrm{MeV}/\mathrm{nucleon}$, indicating more suppression in the medium in the reaction at $56$ $\mathrm{MeV}/\mathrm{nucleon}$. It implies that the in-medium effect depends not only on the local density but also reflects the dynamic situation of the medium, such as a nontrivial phase space distribution. However, previous study in $\mathrm{Sn} + \mathrm{Sn}$ reactions at $270$ $\mathrm{MeV}/\mathrm{nucleon}$ showed that the $\mathrm{AMD}$ calculation cannot simultaneously reproduce the rapidity distributions and the transverse momentum spectra of light charged particles~\cite{SpiRIT:2021och,SpiRIT:2022sqt} by only adjusting parameters of the NN collision matrix element. 


\section{Summary}
In summary, mid-rapidity light charged particles $p$, $d$, $t$, $^3\mathrm{He}$, and $\alpha$ emitted in central ${}^{48}\mathrm{Ca} + {}^{64}\mathrm{Ni}$ and ${}^{40}\mathrm{Ca} + {}^{58}\mathrm{Ni}$ collisions at $56$ and $140$ $\mathrm{MeV}/\mathrm{nucleon}$ were investigated. In both energies, the neutron-rich reactions produce $\sim50-70\%$ more $t$ and $\sim 10-15\%$ more $d$ and $\alpha$ particles, due to the relative abundance of neutrons in the ${}^{48}\mathrm{Ca} + {}^{64}\mathrm{Ni}$ reaction. 

For data comparison, the Antisymmetrized Molecular Dynamics ($\mathrm{AMD}$) model was employed, in which a dynamical description of cluster production is incorporated without relying on coalescence at freeze-out. To ensure a fair comparison between the model and data, events in the $\mathrm{AMD}$ model are selected in the same way as in the experiment based on the charged-particle multiplicity of the Microball coverage.

The screened parameterization, characterized by the screening parameter $\eta$, was adopted in Eq.~\eqref{eq:screening} to describe the reduction of the in-medium nucleon-nucleon matrix element (or $\tilde{\sigma}_\mathrm{NN}$) in $\mathrm{AMD}$. A smaller value of $\eta$ described in Eq.~\eqref{Eq_SigmaNN} implies more reduction of $\tilde{\sigma}_\mathrm{NN}$. 

To study the dependence of particle production on $\tilde{\sigma}_\mathrm{NN}$, the transverse momentum spectra of the light clusters were compared to $\mathrm{AMD}$ calculations. The analysis revealed that transverse momentum spectra at $56$ $\mathrm{MeV}/\mathrm{nucleon}$ were well reproduced with $\eta=0.35$, while spectra at $140$ $\mathrm{MeV}/\mathrm{nucleon}$ were better reproduced with $\eta=0.85$. It is thus useful to obtain data from reactions at different energies to probe the dependence of in-medium effect on the dynamics of the medium. Ultimately, these studies allows us to properly tune in-medium effects in AMD, which is essential for more reliably constraining the symmetry energy using HIC observables.

\section{Acknowledgments}

We would like to acknowledge support from the National Science Foundation Grant Nos. PHY-2110218, PHY-2209145, PHY-1712832 and PHY-2309923. We would also like to acknowledge support from the National Research Foundation of Korea (NRF) grants funded by the Korea government (MSIT) (Nos. 2018R1A5A1025563 and RS-2024-00333673), the JSPS KAKENHI Grant No.~JP21K03528, the U.S. Department of Energy, Office of Science, Nuclear Physics under Award No. DE-SC0021938, and Department of Energy National Nuclear Security Administration Stewardship Science Graduate Fellowship under cooperative Agreement No. DE-NA0002135.


\newpage

\bibliography{ref.bib}
\bibliographystyle{apsrev}

\end{document}